\documentclass[10pt,twocolumn]{article}
\usepackage[a4paper,left=15mm,right=15mm,top=15mm,bottom=15mm,includeheadfoot]{geometry}

\usepackage{mathpazo}
\usepackage{graphicx}
\usepackage{amsmath,textcomp}
\usepackage{makeidx}
\makeindex

\usepackage{ftnright}

\usepackage{color}
\usepackage{graphicx}
\usepackage{hyperref}
\usepackage{pstricks,amssymb}

\usepackage{fancyhdr}

\usepackage[utf8]{inputenc}
\usepackage{geometry}

\usepackage{xcolor}

\usepackage{blindtext}
\usepackage{graphicx}
\usepackage{amsmath}
\usepackage{bm}
\usepackage{amsfonts}
\usepackage{geometry}
\usepackage{amsthm}
\usepackage{gensymb}
\usepackage{authblk}
\title{{On Classical Simulation of Quantum Circuits Composed of Clifford Gates}}
\author{George Biswas \\ georgebsws@gmail.com}
\affil{Department of Physics, Tamkang University, Tamsui Dist., New Taipei 25137, Taiwan, ROC}
\affil{Center for Advanced Quantum Computing, Tamkang University, Tamsui Dist., New Taipei 25137, Taiwan, ROC}
\date{\today}

\begin{document}
\thispagestyle{fancy}

\maketitle

\begin{abstract}
    \textbf{The Gottesman--Knill theorem asserts that quantum circuits composed solely of Clifford gates can be efficiently simulated classically. This theorem hinges on the fact that Clifford gates map Pauli strings to other Pauli strings, thereby allowing for a structured simulation process using classical computations. In this work, we break down the step-by-step procedure of the Gottesman--Knill theorem in a beginner-friendly manner, leveraging concepts such as matrix products, tensor products, commutation, anti-commutation, eigenvalues, and eigenvectors of quantum mechanical operators. Through detailed examples illustrating superposition and entanglement phenomena, we aim to provide a clear understanding of the classical simulation of Clifford gate-based quantum circuits. While we do not provide a formal proof of the theorem, we offer intuitive physical insights at each stage where necessary, empowering readers to grasp the fundamental principles underpinning this intriguing aspect of quantum computation.}
\end{abstract}

\section{Introduction}
The Gottesman--Knill theorem states that quantum circuits containing only Clifford gates can be efficiently simulated classically. This is a well-known concept in the field of quantum computation, which was attributed to Emanuel Knill by D. Gottesman's in his 1997 Ph.D. thesis~\cite[\S~5.7]{gottesman1997stabilizercodesquantumerror}, and published as Knill's theorem in Ref.~\cite{gottesman1998heisenbergrepresentationquantumcomputers} where D.~Gottesman acknowledges private communication with E.~Knill. Further improvement in the theorem and simulation algorithm is also done by S.~Aaronson and D.~Gottesman in Ref.~\cite{PhysRevA.70.052328}.
The complete set of {one- and two-qubit} Clifford gates encompasses the Pauli gates ($X$, $Y$, $Z$), Hadamard gate ($H$), phase gate ($S$), and controlled-not gate ($CNOT$). However, this gate set lacks universality, meaning that it cannot perform all possible quantum computations. Attempt to overcome the universality limitations could be done by the introduction of $T$ gate; However the limitation is there and the formalism is slightly generalized in Ref.~\cite{PhysRevLett.116.250501}.
Classical simulation of quantum circuits involving only Clifford gates is often referred to as the stabilizer formalism \cite[\S~10.5]{Nielsen2010}. A formal proof of the theorem can be found in \cite[\S~10]{Bergou2021}. The purpose of this paper is to help readers to learn the stabilizer formalism and Gottesman--Knill theorem by working through all the detailed intermediate steps using interesting and straightforward examples.
Next, in Sections~\ref{intro_1} and \ref{intro_2}, we will delve into examples of Pauli strings and explore their associated eigenspaces respectively.

\subsection{The Pauli group or the group of Pauli strings}
\label{intro_1}

Single-qubit Pauli strings are represented by $P_1 = {I, X, Y, Z}$, where $I$ is the identity gate, and $X$, $Y$, $Z$ are the Pauli gates \cite{Pauli1927} commonly used in quantum computing \cite[\S~2.1.3]{Nielsen2010}.

For two-qubit Pauli strings, denoted as $P_2$, the set includes all combinations of tensor products of Pauli operators from $P_1$ with themselves. This set is expressed as: $P_2=\{I\otimes I, I\otimes X, I\otimes Y, I\otimes Z, X\otimes X, X\otimes Y, X\otimes Z, X\otimes I, Y\otimes Y, Y\otimes Z, Y\otimes I, Y\otimes X, Z\otimes Z, Z\otimes I, Z\otimes X, Z\otimes Y\}$.

Extending this pattern to $n$ qubits, the $n$-qubit Pauli strings: $P_n=$ {all combinations of} $\{\otimes_{i=1}^n \sigma_i\}$ such that $\sigma_i \in \{I,X,Y,Z\}$. 
Therefore, $P_n$ contains $4^n$ elements, where each element is a tensor product of Pauli gates applied to each qubit in an $n$-qubit system.

The Pauli group $G_n$ associated with $n$ qubits is defined as $G_n = \{P_n, {\pm 1, \pm i}\}$. Here, $P_n$ refers to the set of $n$-qubit Pauli strings and $\{\pm 1, \pm i\}$ represents the set of scalar factors (phases) that can accompany these Pauli strings. This group is complete under multiplication because the product of any two Pauli strings (along with their accompanying phases) results in another Pauli string (potentially multiplied by a phase factor from $\{\pm 1, \pm i\}$).

\subsection{Eigenspace of Pauli strings}
\label{intro_2}

Consider the eigenspace associated with eigenvalue $+1$ of the Pauli string $Z \otimes Z \otimes Z \in G_n$. This eigenspace, denoted as $E_1(Z \otimes Z \otimes Z)$, consists of quantum states $|\psi\rangle \in \mathbb{C}^8$ satisfying the condition $(Z \otimes Z \otimes Z)|\psi\rangle = |\psi\rangle$. 

The basis states of the $E_1(Z \otimes Z \otimes Z)$ eigenspace are defined by the set ${|s_1 s_2 s_3\rangle}$, where each $s_i$ is either $|0\rangle$ or $|1\rangle$. The condition for a state $|\psi\rangle$ to belong to $E_1(Z \otimes Z \otimes Z)$ is that it must have an even number of $|1\rangle$ states among $|s_1\rangle$, $|s_2\rangle$, and $|s_3\rangle$. Conversely, the eigenspace $E_{-1}(Z \otimes Z \otimes Z)$ consists of states where the number of $|1\rangle$ states among $|s_1\rangle$, $|s_2\rangle$, and $|s_3\rangle$ is odd.

In general $E_1(Z^{\otimes n})$ represents quantum states of \{span of $2^{n-1}$ basis states\}.

The Pauli string $Z \otimes Z \otimes Z$ acts as a stabilizer for all quantum states $|\psi\rangle \in E_1(Z \otimes Z \otimes Z)$. This means that any state within this eigenspace will remain unchanged when operated by $Z \otimes Z \otimes Z$.
Note that, in this way, many quantum states can be represented by one stabilizer $Z \otimes Z \otimes Z$. 

This example illustrates the concept of $+1$~eigenspace associated with a Pauli string. Similar scenarios apply to other Pauli strings.

\subsection{Cutting the $E_1$ eigenspace by taking its intersection with another one}\label{intro_3}

In the previous Section~\ref{intro_2}, we learned that a single Pauli string can stabilize multiple quantum states.
The key concept is to leverage the intersection of the $+1$ eigenspaces of multiple Pauli strings to uniquely determine a quantum state.

Consider the example of $X \otimes X$ and $Y \otimes Y$, which commute with each other ($[X \otimes X, Y \otimes Y] = 0$). This commutation property implies that these operators can be simultaneously diagonalized. Specifically, there exists a basis in which both $X \otimes X$ and $Y \otimes Y$ are diagonal. We use computer programming to find that basis in general. However, in this simple example, we see that in the Bell basis, which consists of the states $|\phi^\pm\rangle = \frac{1}{\sqrt{2}}(|00\rangle \pm |11\rangle)$ and $|\psi^\pm\rangle = \frac{1}{\sqrt{2}}(|01\rangle \pm |10\rangle)$, both $X \otimes X$ and $Y \otimes Y$ are diagonal. We find that
\begin{gather*}
(X \otimes X) |\phi^+\rangle = +1 |\phi^+\rangle,\\
(X \otimes X) |\phi^-\rangle = -1 |\phi^-\rangle,\\
(X \otimes X) |\psi^+\rangle = +1 |\psi^+\rangle,\\
(X \otimes X) |\psi^-\rangle = -1 |\psi^-\rangle,\\
(Y \otimes Y) |\phi^+\rangle = -1 |\phi^+\rangle,\\
(Y \otimes Y) |\phi^-\rangle = +1 |\phi^-\rangle,\\
(Y \otimes Y) |\psi^+\rangle = +1 |\psi^+\rangle,\\
(Y \otimes Y) |\psi^-\rangle = -1 |\psi^-\rangle.
\end{gather*}
From this, we identify that the $+1$ eigenspace of $X \otimes X$, denoted $E_1(X \otimes X)$, is spanned by \{$|\phi^+\rangle$ and $|\psi^+\rangle$\}. Similarly, the $+1$ eigenspace of $Y \otimes Y$, denoted $E_1(Y \otimes Y)$, is spanned by \{$|\phi^-\rangle$ and $|\psi^+\rangle$\}.

By taking the intersection of these $+1$ eigenspaces, we find
$$E_1(X \otimes X)\cap E_1(Y \otimes Y)=(|\psi^+\rangle).$$
This intersection represents a unique quantum state, showcasing how intersections of $+1$ eigenspaces of independent commuting stabilizers in the Pauli group can isolate specific quantum states.

In general, if we consider $n$ independent commuting stabilizers from the Pauli group $G_n$ and intersect their $+1$ eigenspaces, the result will yield either one unique quantum state or an empty set (which is not discussed further here).

Thus, through the stabilizer formalism, we can use a set of stabilizers to uniquely represent and characterize individual quantum states.

Next, we will detail the Gottesman--Knill theorem and will explore the theorem with step-by-step examples.

\section{Gottesman Knill theorem}\label{theorem}
Let us break down the process described in the Gottesman--Knill theorem step by step for simulating a quantum circuit using the stabilizer formalism:

\subsection{Initialization with Stabilizers}

The simulation (according to the theorem) begins by starting the quantum circuit with the initial state $|0\rangle^{\otimes n}$, where $n$ is the number of qubits. We aim to represent this state using a set of $n$ independent stabilizers from the Pauli group $G_n$. A typical choice for a set of stabilizers could be $S_n^g = \{(Z\otimes I\otimes I), (I\otimes Z \otimes I), (I\otimes I\otimes Z)\}$ for $n=3$, where each stabilizer has one Pauli $Z$ operator and the rest are $I$ (identity) operators. One important thing to note here is that the set $S_n^g$ is the stabilizer generator set, i.e., the complete set of stabilizers $S_n$ for the quantum state represented by $S_n^g$ can be generated by taking the products of different elements of $S_n^g$.

\subsection{Applying Unitary Operators}

Unitary operations $U$ are applied to the initial state $|0\rangle^{\otimes n}$. The effect of these unitary operators on the stabilizers is crucial. For any stabilizer $g \in S_n^g$ and any state $|\psi\rangle \in E_1(S_n^g)$ (the intersection of $+1$ eigenspace of elements in $S_n^g$), applying $U$ transforms the stabilizer $g$ to $U g U^\dagger$ because we have the following
\begin{align*}
    U|\psi\rangle &=U(g|\psi\rangle), \hspace{0.5cm} \text{Because g is a stabilizer}\\
    &=U g U^{\dagger} U|\psi\rangle, \hspace{0.5cm} \text{Because}\hspace{0.1cm} U^{\dagger}U=I\\
    &=(U g U^{\dagger}) U|\psi\rangle.
\end{align*} 
This transformation updates the set of stabilizers to $US_n^g U^\dagger = \{Ug_1U^\dagger, ..., Ug_nU^\dagger\}$,

In the stabilizer formalism, we have the restriction that the elements of $US_n^g U^{\dagger}$ must belong to the Pauli Group $G_n$; That is, only those unitary operators are allowed that map a Pauli string to another Pauli string. The Clifford gates are those allowed unitaries. Therefore, in the context of classical simulation of quantum circuits, Clifford gates are those one- and two-qubit gates that map any Pauli string to another Pauli string. Here, we focus on the simplest Clifford algebra, also known as the \emph{Pauli algebra} \cite{Marsh2018}. Higher-dimensional (general) Clifford algebras also exist and are discussed in the literature~\cite{Quanta199,PhysRevA.63.054302}.
Any quantum gate can be represented in matrix form. The matrix forms of the Clifford gates in the spin-$z$ basis $\left\{ |\uparrow_{z}\rangle,|\downarrow_{z}\rangle\right\}$ or the computational basis $\left\{ |0\rangle,|1\rangle\right\}$ are as follows
\begin{gather*}
\text{X} = \begin{pmatrix}
0 & 1 \\
1 & 0
\end{pmatrix}
, \\
\text{Y} = \begin{pmatrix}
0 & -i \\
i & 0
\end{pmatrix}
,\\
\text{Z} = \begin{pmatrix}
1 & 0 \\
0 & -1
\end{pmatrix}
,\\
\text{H} = \frac{1}{\sqrt{2}}\begin{pmatrix}
1 & 1 \\
1 & -1
\end{pmatrix}
,\\
\text{S} = \begin{pmatrix}
1 & 0 \\
0 & i
\end{pmatrix}
,\\
\text{CNOT} = \begin{pmatrix}
1 & 0 & 0 & 0 \\
0 & 1 & 0 & 0 \\
0 & 0 & 0 & 1 \\
0 & 0 & 1 & 0
\end{pmatrix}.   
\end{gather*}
The action of these gates on the set of stabilizers is constructed with matrix multiplication in the form of so-called \emph{adjoint action}, where stabilizers are enclosed between considered matrix and its inverse (Hermitian conjugate).

\subsection{Measurement in the \texorpdfstring{$Z^{\otimes n}$}{Z\textasciicircum n} Basis}

To measure the quantum state in the computational basis ($Z^{\otimes n}$ basis), each qubit is measured individually. For instance, to measure qubit 1 in a three-qubit system, we use the measurement operator $(Z\otimes I\otimes I)$. In the classical simulation, we check if this measurement operator commutes with all the stabilizer generators in the updated set $US_n^g U^\dagger$:

Case~1: If the measurement observable commutes with all stabilizer generators, it indicates a definite measurement outcome. However, to determine if the outcome is $+1$ or $-1$, we may need to consider if the measurement observable itself is a stabilizer or if its negative counterpart is a stabilizer. No change occurs in the set of stabilizers in this case.

Case~2: If the measurement observable anti-commutes with at least one stabilizer generator, this signifies a probabilistic measurement outcome (with equal probability of $+1$ and $-1$). The anti-commuting stabilizer generator is then replaced by ($\pm$ measurement operator) based on the measurement result ($+1$ or $-1$).\\

In summary, the Gottesman--Knill theorem utilizes the stabilizer formalism to efficiently simulate quantum circuits composed of Clifford gates by representing quantum states with stabilizers and updating these stabilizers through unitary operations and measurements.

Next, in Section~\ref{E_1}, we will work through an example of simulating a 3-qubit quantum circuit using single-qubit Clifford gates to further elucidate the application of the theorem.

\section{Three-qubit quantum circuit}
\label{E_1}

\begin{figure}[h!]
    \centering
    \includegraphics[width=\linewidth]{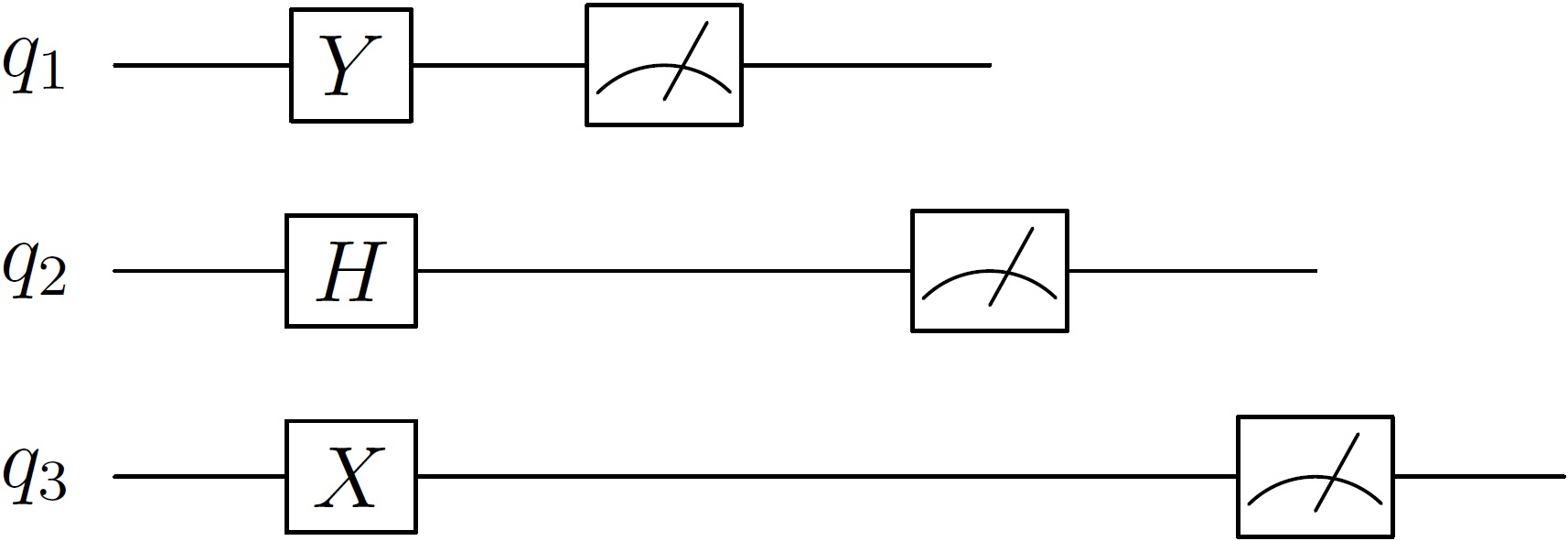}
    \caption{Diagram of the $3$-qubit circuit, we want to simulate.}
    \label{e1}
\end{figure}

Let us simulate a 3-qubit quantum circuit starting with the initial state $|000\rangle$ using the stabilizer formalism. We begin by applying specific quantum gates to each qubit: the Pauli Y gate ($Y$) to qubit 1, the Hadamard gate ($H$) to qubit 2, and the Pauli X gate ($X$) to qubit 3. To accomplish this simulation, we update the stabilizers of the initial state to incorporate the effects of these gates on each qubit. Subsequently, we proceed to measure each qubit individually on the computational basis (Z basis). The quantum circuit of the considered 3-qubit example is depicted in Fig.~\ref{e1}.

Let us work out step by step:

Step 1: Initialization:
We start with the initial quantum state $|0\rangle^{\otimes 3}$, which can be represented by a set of three independent stabilizer generators as follows
\begin{align}
\label{SG_initial_3}
    SG_\textrm{initial}=\{(Z\otimes I\otimes I), (I\otimes Z \otimes I),
    (I\otimes I\otimes Z)\}.
\end{align}

Step 2: Applying gates:
Next, we apply specific gates to each qubit. 
The action of these gates on the stabilizer generator set $SG_\textrm{initial}$ is represented by the updated stabilizer generator set $SG_\textrm{final}$:
\begin{align}
\label{SG_after_first_unitary_layer_3} 
    SG_\textrm{final}&=\{(YZY^{\dagger} \otimes HIH^{\dagger} \otimes XIX^{\dagger}), \notag \\ \notag & \hspace{0.5cm} (YIY^{\dagger} \otimes HZH^{\dagger} \otimes XIX^{\dagger}),\\ \notag &  \hspace{5mm} (YIY^{\dagger} \otimes HIH^{\dagger} \otimes XZX^{\dagger})\}\\ 
    &=\{(- Z\otimes I\otimes I), (I\otimes X \otimes I),  (- I\otimes I\otimes Z)\} 
.
\end{align}
Therefore, after applying this layer of Clifford gates, the stabilizer generator set $SG_\textrm{final}$ represents the updated state of the quantum system, incorporating the transformations induced by the gate operations on each qubit within the stabilizer formalism.\\

Step 3: Measurement of qubit 1:
To measure qubit 1 in the computational (or Z) basis, we use the measurement operator $(Z \otimes I \otimes I)$. This operator commutes with all stabilizer generators in the final set $SG_\textrm{final}$ obtained after applying gates to the initial state. Because $(Z \otimes I \otimes I)$ commutes with all stabilizer generators, the measurement outcome for qubit 1 is deterministic. However, to determine the eigenvalue ($+1$ or $-1$) of qubit 1, we check if $(Z \otimes I \otimes I)$ or $(-Z \otimes I \otimes I)$ is a stabilizer of the quantum state represented by $SG_\textrm{final}$. In this case, $(-Z \otimes I \otimes I)$ is a stabilizer in $SG_\textrm{final}$, indicating that the eigenvalue of qubit 1 is $-1$.

Because the quantum state is an eigenvector of the measurement operator, it will not change the post-measurement state, and we do not have to make any update to the stabilizer generator set.\\

Step 4: Measurement of qubit 2:
The measurement operator for qubit 2 is $(I \otimes Z \otimes I)$. This operator anti-commutes with the stabilizer generator $(I \otimes X \otimes I)$ in $SG_\textrm{final}$, leading to a probabilistic measurement outcome with equal probability of $+1$ and $-1$. We use a random number generator (uniform random $\pm 1$ generator) to determine the outcome. Depending on the measurement result ($+1$ or $-1$), we update the stabilizer generator set $SG_\textrm{final}$ accordingly. If the outcome is $+1$, $SG_{q_2=+1}$ is formed; if the outcome is $-1$, $SG_{q_2=-1}$ is formed. 
\begin{gather*}
SG_{q_2=+1}=\{(- Z\otimes I\otimes I), (I\otimes Z \otimes I), (- I\otimes I\otimes Z)\},  \\
SG_{q_2=-1}=\{(- Z\otimes I\otimes I), (- I\otimes Z \otimes I),  (- I\otimes I\otimes Z)\}.
\end{gather*}

Step 5: Measurement of qubit 3:
For qubit 3, the measurement operator is $(I \otimes I \otimes Z)$. We use the updated stabilizer generator set $SG_{q_2=+1}$ or $SG_{q_2=-1}$ obtained from the previous step. This measurement operator commutes with all stabilizer generators in the corresponding set, ensuring a deterministic measurement outcome. In this case, ${(- I \otimes I \otimes Z)}$ is a stabilizer, indicating that the eigenvalue of qubit 3 is $-1$.

Because the quantum state is an eigenvector of the measurement operator,
there is no update in the set of stabilizer generators.\\

Final Measurement Results:
Note that a measurement outcome of $+1$ corresponds to $|0\rangle$ and a measurement outcome of $-1$ corresponds to $|1\rangle$ because $|0\rangle$ is an eigenstate of $Z$ with eigenvalue $+1$ and $|1\rangle$ is an eigenstate of $Z$ with eigenvalue $-1$.

Based on the measurement outcomes, we have $q_1 = -1$ (corresponding to $|1\rangle$), $q_2 = \pm 1$ (corresponding to $|0\rangle$ or $|1\rangle$), and $q_3 = -1$ ( corresponding to $|1\rangle$). Therefore, the final measurement results represent a superposition of states $|101\rangle$ and $|111\rangle$.\\

In this simplest example with a single layer of single-qubit gates, we observe quantum superposition phenomena and apply the Gottesman--Knill theorem step by step to simulate the quantum circuit using stabilizer formalism. For a more comprehensive illustration, we will work through an example with two layers of Clifford gates in next Section~\ref{E_2} to demonstrate entanglement phenomena and further understand the application of the theorem.

\section{Four-qubit quantum circuit}
\label{E_2}

\begin{figure}[h!]
    \centering
    \includegraphics[width=\linewidth]{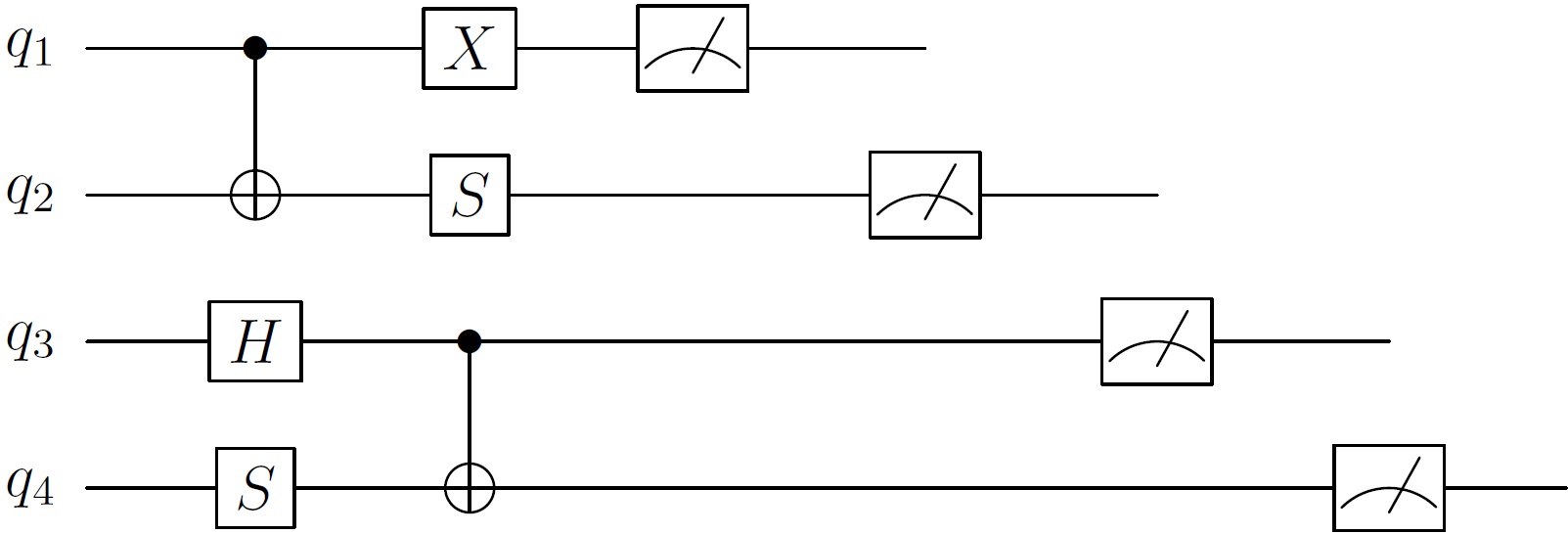}
    \caption{Diagram of the $4$-qubit circuit, we want to simulate.}
    \label{e2}
\end{figure}

Let us analyze a quantum circuit operating on a 4-qubit system, composed of two layers of operations. In the first layer, a CNOT gate is applied with qubit 1 as the control and qubit 2 as the target. Additionally, a Hadamard gate is applied to qubit 3, and an S gate (also known as the phase gate) is applied to qubit 4. This layer prepares qubit 3 in a superposition state via the Hadamard gate, and applies a phase shift to qubit 4 using the S gate. In the second layer, a Pauli X gate (NOT gate) is applied to qubit 1, an S gate is applied to qubit 2, and another CNOT gate is applied with qubit 3 as the control and qubit 4 as the target. This subsequent layer further manipulates the quantum state, flipping the state of qubit 1, adjusting the phase of qubit 2, and potentially entangling qubits 3 and 4 based on the CNOT gate operation. This multi-layered quantum circuit demonstrates the sequential application of gates to manipulate the state of a multi-qubit system, showcasing fundamental quantum operations and potential entanglement effects. Fig.~\ref{e2} depicts the 4-qubit quantum circuit.\\

Step 1: Represent the initial state $|0\rangle^{\otimes 4}$:
The initial set of stabilizer generators can be written as
\begin{align}
\label{SG_initial_4}
SG_\textrm{initial}
&=\{(Z\otimes I\otimes I \otimes I), \notag \\ 
& \hspace{0.5cm}(I\otimes Z \otimes I \otimes I), \notag \\ 
& \hspace{0.5cm} (I\otimes I\otimes Z \otimes I), \notag \\ 
& \hspace{0.5cm} (I\otimes I\otimes I \otimes Z)\}
.
\end{align}

Step 2: Apply the first layer of Clifford gates:
Applying the gates the set of stabilizer generators is modified to $SG_\textrm{mid}$ as
\begin{align}
\label{SG_intermediate_4}
SG_\textrm{mid}
&=\{(CNOT_{1,2}\ Z \otimes I \ CNOT_{1,2}^{\dagger} \otimes H I H^{\dagger} \otimes S I S^{\dagger}), \nonumber \\
&\hspace{5mm} (CNOT_{1,2}\ I\otimes Z \ CNOT_{1,2}^{\dagger} \otimes  H I H^{\dagger} \otimes S I S^{\dagger}), \nonumber\\
&\hspace{5mm} (CNOT_{1,2}\ I\otimes I \ CNOT_{1,2}^{\dagger} \otimes  H Z H^{\dagger} \otimes S I S^{\dagger}), \nonumber\\
&\hspace{5mm} (CNOT_{1,2}\ I\otimes I \ CNOT_{1,2}^{\dagger} \otimes  H I H^{\dagger} \otimes S Z S^{\dagger})\} \nonumber\\
&=\{(Z\otimes I\otimes I \otimes I), (Z\otimes Z \otimes I \otimes I), \nonumber \\
& \hspace{0.5cm} (I\otimes I\otimes X \otimes I),  (I\otimes I\otimes I \otimes Z)\}
.
\end{align}

Step 3: Apply the second layer of Clifford gates: Applying the gates we get the modified stabilizer generators $SG_\textrm{final}$ as
\begin{align}
\label{SG_final_4}
SG_\textrm{final}
&=\{(X Z X^{\dagger} \otimes S I S^{\dagger} \otimes CNOT_{3,4}\ I \otimes I \ CNOT_{3,4}^{\dagger}), \nonumber\\
&\hspace{5mm} (X Z X^{\dagger} \otimes S Z S^{\dagger} \otimes CNOT_{3,4}\ I \otimes I \ CNOT_{3,4}^{\dagger}), \nonumber\\
&\hspace{5mm} (X I X^{\dagger} \otimes S I S^{\dagger} \otimes CNOT_{3,4}\ X \otimes I \ CNOT_{3,4}^{\dagger}), \nonumber\\
&\hspace{5mm} (X I X^{\dagger} \otimes S I S^{\dagger} \otimes CNOT_{3,4}\ I \otimes Z \ CNOT_{3,4}^{\dagger})\} \nonumber\\
&=\{(- Z\otimes I\otimes I \otimes I), (- Z\otimes Z \otimes I \otimes I), \nonumber \\
& \hspace{0.5cm} (I\otimes I\otimes X \otimes X), (I\otimes I\otimes Z \otimes Z)\}
.
\end{align}

Step 4: Measure qubit~1: To measure qubit 1, we use the measurement operator $(Z \otimes I \otimes I \otimes I)$. This operator commutes with all stabilizer generators in the final set $SG_\textrm{final}$, and $(-Z \otimes I \otimes I \otimes I)$ is a stabilizer in $SG_\textrm{final}$. Therefore, the measurement outcome for qubit 1 is $-1$ (corresponding to the state $|1\rangle$), and there is no change in the stabilizer generators post-measurement.\\

Step 5: Measure qubit~2: The measurement operator for qubit 2 is $(I \otimes Z \otimes I \otimes I)$. This operator commutes with all stabilizer generators in $SG_\textrm{final}$. Although $(I \otimes Z \otimes I \otimes I)$ is not explicitly in $SG_\textrm{final}$, it can be found in the corresponding set of stabilizers $S_\textrm{final}$ (because $(-Z \otimes I \otimes I \otimes I)(-Z \otimes Z \otimes I \otimes I) = (I \otimes Z \otimes I \otimes I)$). The measurement outcome for qubit 2 is $+1$, and there is no change in the stabilizer generators post-measurement.\\

Step 6: Measure qubit~3: For qubit 3, the measurement operator is $(I \otimes I \otimes Z \otimes I)$. This operator anti-commutes with the stabilizer generator $(I \otimes I \otimes X \otimes X)$ in $SG_\textrm{final}$. Therefore, the measurement outcome for qubit 3 is probabilistic (with equal probability of $\pm 1$), leading to two possible updated sets of stabilizer generators based on the measurement result:

Case~1 (measurement result $+1$): Updated stabilizer generators $SG_{q_3=+1}$.

$SG_{q_3=+1}  =\{(- Z\otimes I\otimes I \otimes I), (- Z\otimes Z \otimes I \otimes I),$

\hspace{1.8cm} $(I\otimes I\otimes Z \otimes I), (I\otimes I\otimes Z \otimes Z)\}.$

Case~2 (measurement result $-1$): Updated stabilizer generators $SG_{q_3=-1}$.

$SG_{q_3=-1}=\{(- Z\otimes I\otimes I \otimes I), (- Z\otimes Z \otimes I \otimes I),$

\hspace{1.8cm} $(- I\otimes I\otimes Z \otimes I), (I\otimes I\otimes Z \otimes Z)\}$.\\

Step 7: Measure qubit~4: Using the updated stabilizer generator set from the previous step, qubit~4 is measured with the operator $(I \otimes I \otimes I \otimes Z)$. This operator commutes with all stabilizers in both $SG_{q_3=+1}$ and $SG_{q_3=-1}$. The measurement outcome for qubit~4 is deterministic and depends on the result of qubit~3's measurement:

If qubit~3's measurement outcome was $+1$, the measurement outcome for qubit~4 is $+1$.

If qubit 3's measurement outcome was $-1$, the measurement outcome for qubit 4 is $-1$.

Either the measurement outcomes of both of them are $+1$ or both of them are $-1$, revealing entanglement between the two qubits.

Therefore, observing the measurement results ($q_1 = -1$, $q_2 = +1$, $q_3 = \pm 1$, and $q_4 = +1$ or $q_4 = -1$ depending on the measurement result of $q_3$), we conclude that the final measurement results represent a superposition of states $|1000\rangle$ and $|1011\rangle$. This analysis demonstrates entanglement between qubit 3 and qubit 4, highlighting the complex interactions and correlations that arise in multi-qubit quantum systems.

\section{Discussion}

Several issues may deserve some further elaboration.

First, in the case of probabilistic measurement outcomes where the measurement operator anti-commutes with a stabilizer generator, resulting in equal probabilities of $\pm 1$, the underlying reason can be understood through the properties of Pauli matrices and their eigenvectors. For instance, the eigenvectors of the Pauli $X$ operator are $\frac{|0\rangle \pm |1\rangle}{\sqrt{2}}$, and when measured in the computational basis (Pauli $Z$ basis), they yield equal probabilities of obtaining $|0\rangle$ and $|1\rangle$. This principle extends to other Pauli matrices like $Y$, where measurement in the computational basis also results in equal probabilities of $\pm 1$. If Pauli $X$ or $Y$ is a stabilizer of a quantum state then the quantum state is an eigenvector of Pauli $X$ or Pauli $Y$ operator, and Pauli $X$ or $Y$ operator anti-commutes with Pauli $Z$ operator. Therefore, if the measurement operator anti-commutes with stabilizer we have equal probability of getting $|0\rangle$ and $|1\rangle$. In the 3-qubit example, Step~4, the quantum state was an eigenvector of $(I \otimes X \otimes I)$ and we measure that eigenvector in the basis of $(I \otimes Z \otimes I)$. The state was in an equal superposition of measurement basis. Therefore, we get outcome $\pm 1$ with equal probability.

Second, we update the anti-commuting stabilizer generator with the measurement operator in the case of probabilistic measurement. If more than one stabilizer generators anti-commutes with the measurement operator, their product will commute with the measurement operator. For instance, if two stabilizer generators like $(I \otimes X \otimes X)$ and $(I \otimes X \otimes I)$ both anti-commute with $(I \otimes Z \otimes I)$, their product $(I \otimes X \otimes X)(I \otimes X \otimes I)$ will commute with $(I \otimes Z \otimes I)$ and can be used to update the stabilizer set. Therefore, we update one stabilizer generator with the measurement observable and another stabilizer generator with the product of the two anti-commuting stabilizer generators. Similarly, if we have more anti-commuting stabilizer generators, we update them by their product with the stabilizer generator we initially replaced by the measurement operator.

Third, the exponential growth in memory required to store the tensor product of $n$ ($2 \times 2$) Pauli matrices poses a challenge for classical computation. However, a clever trick to overcome this issue is to encode Pauli strings using only 2 bits per Pauli matrix. For instance, using a mapping like $00$ for $I$, $01$ for $X$, $10$ for $Y$, and $11$ for $Z$, an array of $n$ Pauli matrices can be encoded in $2n + 1$ bits of memory (one extra bit is required to encode the sign ($\pm$) of Pauli strings), which is of polynomial order in terms of $n$. This efficient encoding drastically reduces the memory requirements for representing and manipulating Pauli strings.

Fourth, applying Clifford gates to encoded Pauli strings represented by arrays of bits does not require direct matrix multiplication. Instead, we define a set of rules to update individual array elements based on the same results achieved through matrix multiplications. These rules are based on the action of the Clifford gates on the Pauli matrices and their encoded representations. 
 
The trick is that there are only a few of those rules.
We may list them down:
\begin{gather*}
HXH^{\dagger}=Z,\\
 HYH^{\dagger}=-Y,\\
 HZH^{\dagger}=X, \\
SXS^{\dagger}=Y,\\
 SYS^{\dagger}=-X,\\
 SZS^{\dagger}=Z, \\
XXX^{\dagger}=X,\\
 XYX^{\dagger}=-Y,\\
 XZX^{\dagger}=-Z, \\
YXY^{\dagger}=-X,\\
 YYY^{\dagger}=Y,\\
 YZY^{\dagger}=-Z,\\
ZXZ^{\dagger}=-X,\\
 ZYZ^{\dagger}=-Y,\\
 ZZZ^{\dagger}=Z,\\
CNOT_{1,2}\ I_1 \otimes I_2 \ CNOT_{1,2}^{\dagger}=I_1 \otimes I_2, \\
CNOT_{1,2}\ I_1 \otimes Z_2 \ CNOT_{1,2}^{\dagger}=Z_1 \otimes Z_2, \\
CNOT_{1,2}\ Z_1 \otimes I_2 \ CNOT_{1,2}^{\dagger}=Z_1 \otimes I_2, \\
CNOT_{1,2}\ Z_1 \otimes Z_2 \ CNOT_{1,2}^{\dagger}=I_1 \otimes Z_2, \\
CNOT_{1,2}\ I_1 \otimes X_2 \ CNOT_{1,2}^{\dagger}=I_1 \otimes X_2, \\
CNOT_{1,2}\ X_1 \otimes I_2 \ CNOT_{1,2}^{\dagger}=X_1 \otimes X_2, \\
CNOT_{1,2}\ X_1 \otimes X_2 \ CNOT_{1,2}^{\dagger}=X_1 \otimes I_2, \\
CNOT_{1,2}\ Z_1 \otimes X_2 \ CNOT_{1,2}^{\dagger}=Z_1 \otimes X_2, \\
CNOT_{1,2}\ X_1 \otimes Z_2 \ CNOT_{1,2}^{\dagger}=- Y_1 \otimes Y_2, \\
CNOT_{1,2}\ I_1 \otimes Y_2 \ CNOT_{1,2}^{\dagger}= Z_1 \otimes Y_2, \\
CNOT_{1,2}\ Y_1 \otimes I_2 \ CNOT_{1,2}^{\dagger}= Y_1 \otimes X_2, \\
CNOT_{1,2}\ X_1 \otimes Y_2 \ CNOT_{1,2}^{\dagger}= Y_1 \otimes Z_2, \\
CNOT_{1,2}\ Y_1 \otimes X_2 \ CNOT_{1,2}^{\dagger}= Y_1 \otimes I_2, \\
CNOT_{1,2}\ Z_1 \otimes Y_2 \ CNOT_{1,2}^{\dagger}= I_1 \otimes Y_2, \\
CNOT_{1,2}\ Y_1 \otimes Z_2 \ CNOT_{1,2}^{\dagger}= X_1 \otimes Y_2, \\
CNOT_{1,2}\ Y_1 \otimes Y_2 \ CNOT_{1,2}^{\dagger}=- X_1 \otimes Z_2,
\end{gather*}
where for $CNOT_{1,2}$ qubit~1 is control and qubit~2 is the target. 

By applying these rules systematically, we can efficiently simulate the action of Clifford gates on encoded Pauli strings without the need for direct matrix operations, thereby streamlining classical computations involving quantum states and operations.

\section{Significance}

The stabilizer formalism is a powerful framework in quantum computing that greatly simplifies the description and analysis of quantum error-correcting codes and quantum algorithms involving Clifford gates. The Gottesman--Knill theorem is pivotal in delineating the boundary between quantum computations that can be efficiently simulated classically (hence do not provide quantum advantage) and those that provide a genuine quantum advantage. This understanding informs the development of more complex quantum algorithms and error-correction methods in contemporary quantum computing research.

Recent advancements have refined techniques for efficiently simulating larger quantum systems using the stabilizer formalism. New algorithms and methods have been introduced to extend these simulations to more complex circuits and enhance error-correcting codes~\cite{Gidney2021stimfaststabilizer, PhysRevA.73.022334, Gosset2024fastsimulationof}. Additionally, experimental implementations of stabilizer-based error correction schemes have been successfully demonstrated~\cite{Livingston2022}. The extended stabilizer formalism is also actively employed to benchmark the performance of quantum computers~\cite{PhysRevLett.129.150502, PhysRevA.108.052613, Govia_2023}.

Future research could focus on extending the stabilizer formalism to accommodate more general quantum gates and exploring hybrid classical-quantum algorithms that leverage both stabilizer methods and quantum advantages. These developments have the potential to offer new avenues for enhancing computational efficiency and advancing error mitigation strategies.

\section*{Acknowledgement}
The author thanks Alexander Vlasov and Danko D. Georgiev (editors, Quanta Journal) for their valuable corrections and modifications to this published version.

\vspace{0.5cm}

Journal Ref: Quanta 2024; 13: 20-27.

DOI: 10.12743/quanta.v13i1.265

\bibliographystyle{ieeetr}
\bibliography{ref.bib}
\end{document}